\documentclass[acmsmall,screen,nonacm,review=false]{acmart}

\setcopyright{none}

\usepackage{wrapfig}
\usepackage{subcaption}
\usepackage{tikz}
\usepackage{graphicx}
\usepackage{natbib}
\usepackage{pifont}
\usepackage{nicematrix}

\newcommand{\cmark}{\ding{51}}
\newcommand{\xmark}{\ding{55}}

\title{Making Hybrid Languages: A Recipe}

\author{Leif Andersen}
\email{Leif.Andersen@umb.edu}
\affiliation{\institution{University of Massachusetts Boston}
\city{Boston}
\state{Massachusetts}
\country{USA}}
\author{Cameron Moy}
\email{camoy@ccs.neu.edu}
\affiliation{\institution{Northeastern University}
\city{Boston}
\state{Massachusetts}
\country{USA}}
\author{Stephen Chang}
\email{Stephen.Chang@umb.edu}
\affiliation{\institution{University of Massachusetts Boston}
\city{Boston}
\state{Massachusetts}
\country{USA}}
\author{Matthias Felleisen}
\email{matthias@ccs.neu.edu}
\affiliation{\institution{Northeastern University}
\city{Boston}
\state{Massachusetts}
\country{USA}}

\begin{abstract}

  The dominant programming languages support only linear text to express ideas.
  Visual languages offer graphical representations for entire programs,
  when viewed with special tools.
  Hybrid languages, with support from existing
  tools, allow developers to express their ideas with a mix of textual and
  graphical syntax tailored to an application domain. This mix puts both kinds
  of syntax on equal footing and, importantly, the enriched language does not
  disrupt a programmer's typical workflow. This paper presents a recipe for
  equipping existing textual programming languages as well as accompanying IDEs
  with a mechanism for creating and using graphical interactive syntax. It also
  presents the first hybrid language and IDE created using the recipe.

\end{abstract}

\begin{document}
\maketitle

\section{Mixing Text with Visual and Interactive Syntax} \label{sec:intro}

Programmers use programming languages to communicate their thoughts, both to
computers and to other programmers. Linear text suffices for this purpose most
of the time, but some thoughts are inherently geometric and better expressed
visually.

Recognizing this problem, researchers have devised many solutions ranging from
purely visual languages~\cite{rmmrebmrssk:scratch}, to special-purpose
IDEs~\cite{pg:ipython}, and
\begin{wrapfigure}{r}{0.40\textwidth} \centering
  \includegraphics[scale=0.8]{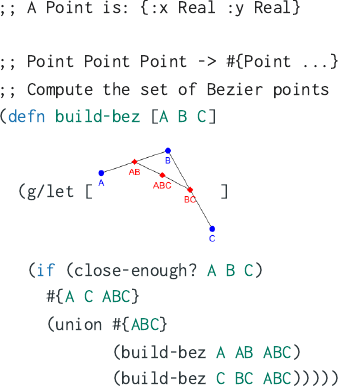}
\caption{Computing Bézier points}\label{fig:bezier}
\end{wrapfigure}
various other strategies~\cite{sandblocks,
ocmvc:livelits}. In particular, \citet{abf:adding} proposed the idea of a
\emph{hybrid language}, which combines textual code with
miniature graphical user interfaces (GUIs), dubbed visual and interactive
syntax. Using a hybrid language, programmers can communicate their thoughts with
text most of the time and weave in visual interactive constructs when
it is appropriate for the problem domain.

For example, consider the calculation of quadratic Bézier curves~\cite{f:curves}.
The standard algorithm combines two tasks: finding midpoints and recursion.
While the second is easily expressed via text, the first is a geometric idea
that deserves a pictorial representation. The {\tt build-bez} function in
figure~\ref{fig:bezier} illustrates how a programmer might use a hybrid variant
of Clojure to convey these two ideas. It computes a set of Bézier points from
three input points that form a triangle. While the embedded visual syntax
depicts the midpoint (of midpoints) calculation, recursively computing the rest
of the points remains textual.

Unfortunately, existing attempts at visual syntax have fundamental
flaws that either unreasonably disrupt a programmer's typical
workflow, impose an undue burden on a language implementation, or
both. For example, such solutions almost always force programmers to
use one particular (new) IDE for their work, a non-starter for
most. \citet{abf:adding}'s solution forces a language implementation
to maintain {\em two\/} different GUI libraries and keep them
synchronized---when even one such library already imposes a serious
amount of work for most languages. This duplication makes the
implementation costly to unmaintain. It also means programmers will
often have to implement the same GUI twice: once for the actual user
interface---using the original GUI library---and a second time for the
visual syntax renderer---using the special-purpose library.

This paper's main contribution is the development of a general recipe for
creating {\em maintainable} and {\em usable} hybrid languages. All prior work
lacks at least one of these attributes. A key idea is that hybrid languages
should be created by adapting existing languages and IDEs instead of creating
new ones. Doing so improves usability because it allows programmers to keep
using a familiar language, and a popular IDE, when programming with the hybrid
language. It also helps maintainability because it reuses a language's existing
infrastructure and libraries. To show this concretely, a second contribution of
this paper is to apply the recipe to create \textsc{Hybrid Clojure\-Script} and a compatible
CodeMirror-based hybrid IDE.

More specifically, section~\ref{sec:design-space} describes the design goals of
the recipe, which come from studying existing solutions. Next,
section~\ref{sec:ingredients} presents the ingredients needed for the recipe: an
existing programming language, a general-purpose IDE, and a GUI library.
Section~\ref{sec:preview} provides a quick glimpse at the result of applying the
recipe to ClojureScript and other chosen ingredients. The subsequent two
sections explain the recipe in detail. Specifically,
section~\ref{sec:design-lang} explains how to adapt a language to support hybrid
syntax, and section~\ref{sec:design-ide} explains how to adapt an existing IDE
so that it may visualize hybrid syntax and allow programmers to interact
with it. Next, section~\ref{sec:evaluation} compares various hybrid languages and
IDEs along several dimensions, and section~\ref{sec:examples} describes several
case studies. Finally, the last two sections explain related work and conclude.

\section{Requirements for Creating a Hybrid Language} \label{sec:design-space}

An examination of existing work suggests the following guidelines for a hybrid-language recipe:
\begin{enumerate}

\item The goal must be to adapt an existing language. Doing so immediately makes
 the resulting language usable because programmers can continue working in a
 familiar context.  The result is maintainable because
 implementers do not need to work on new system.

\item Likewise, hybrid IDEs should be adapted from existing ones. This comes with
 the same advantages for IDEs as for the languages.

\item The hybrid language must be backwards compatible, meaning that it can run
existing non-hybrid code, and that hybrid programs can run on non-hybrid
implementations.

\item All existing IDEs and text editors must work with hybrid code.
 Since programmers have strong IDE preferences, forcing a specific IDE choice
 imposes a serious usability burden on programmers. Hybrid code should also
 remain compatible with other existing tools.

\item Implementations of visual syntax must be able to re-use existing GUIs and
 GUI libraries. This relieves language implementers from having to maintain two
 different GUI libraries. It also helps programmers avoid duplicate effort,
 because they will often create interactive-syntax extensions that are identical
 to the GUI found in the application's run-time code.

\item Finally, visual and interactive syntax should be linguistic,
 meaning it must smoothly integrate with textual syntax and all the
 language's abstraction mechanisms. Similarly, visual syntax should
 not just be a new category of syntax, but it should ideally be
 possible to turn all existing syntactic categories of the
 chosen language into visual syntax constructs.

\end{enumerate}
Section~\ref{sec:related} presents a more detailed analysis of the most closely related
pieces of research and how each of them lives up to the above guidelines.


\section{The Ingredients} \label{sec:ingredients}

Every recipe starts with a list of ingredients. As described in the last
section, the basic ingredients for making a hybrid language are (1) an existing
programming language; (2) an IDE, and (3) a GUI library. Since different quality
ingredients can affect the outcome of a recipe, this section describes some
additional attributes that facilitates the construction of hybrid languages and
IDEs, and also makes the results truly usable.

\subsection{Selecting High-Quality Ingredients}

Ideally, the chosen language comes with a syntax-extension mechanism. Since the
goal is to add interactive syntax for any problem domain, a programmer needs a
mechanism for interpreting new syntactic features in the language.  While it is
possible to create shallow embeddings of domain-specific notations in any
language, a syntactically extensible language greatly facilitates this step.

Similarly, the chosen IDE should (1) provide an extension interface for plug-ins
and (2) support the execution of code at edit time. By using a plug-in tool, it
becomes possible to interpret visual syntax extensions as mini-GUIs in the IDE's
editor. Since this interpretation must run while the programmer edits code, an
IDE that can run code at edit time in an isolated fashion is the best match.
After all, programmer-created code may accidentally interfere with a logical invariant
of the IDE's implementation if it is run without protection. Of course, this
edit-time code must simultaneously be written in the chosen language and must
cooperate with the IDE's editor---which suggests additional constraints on the
chosen GUI library.

Besides being suitable for building application-level graphical interfaces, the
GUI library must come with a text editor that is the same as the chosen IDE's
editor. Furthermore, the editor must allow the insertion of GUI widgets
(canvases, buttons, menus). By meeting these two criteria, programmers should
easily be able to share run-time GUI code with edit-time GUI syntax extensions.


\subsection{Example Ingredients}

Given these attributes, ClojureScript, CodeMirror, and the DOM (Document Object
Model) are reasonable choices. ClojureScript supplies a Lisp-style macro system
that makes it easy to create a construct for defining visual and interactive
syntax. The CodeMirror IDE comes with a rich plugin API, though it does not
inherently isolate code that runs at edit time.

As for the GUI attribute, ClojureScript is a scripting language for the DOM, and
Code\-Mirror's editor is based on the DOM as well. These ingredients can
smoothly collaborate at edit time, as long as the IDE can be protected from
problems in the programmer-supplied code for interactive syntax extensions.

Of the three ingredients, the DOM is particularly beneficial. It is standardized
and can be found at the heart of web browsers, modern IDEs such as Visual Studio
Code, and even some native operating systems like Android. Three decades of
development work have turned it into a performant and expressive technology. At
the same time, programmers have contributed a large collection of widely
available GUI libraries for the DOM such as React, Angular, and Vue.js. All of
these are familiar to many programmers and can thus be used to quickly build
both application-level GUIs and re-used for the creation of interactive syntax
extensions.




\section{A First Taste}\label{sec:preview}

Tasting a dish is often incentive to ask for the recipe. In this spirit, this
section provides a brief overview of \textsc{Hybrid Clojure\-Script} and its use in \texttt{elIDE},
the hybrid CodeMirror-based IDE.

\begin{figure}[t]
  \centering
  \includegraphics[width=.9\textwidth]{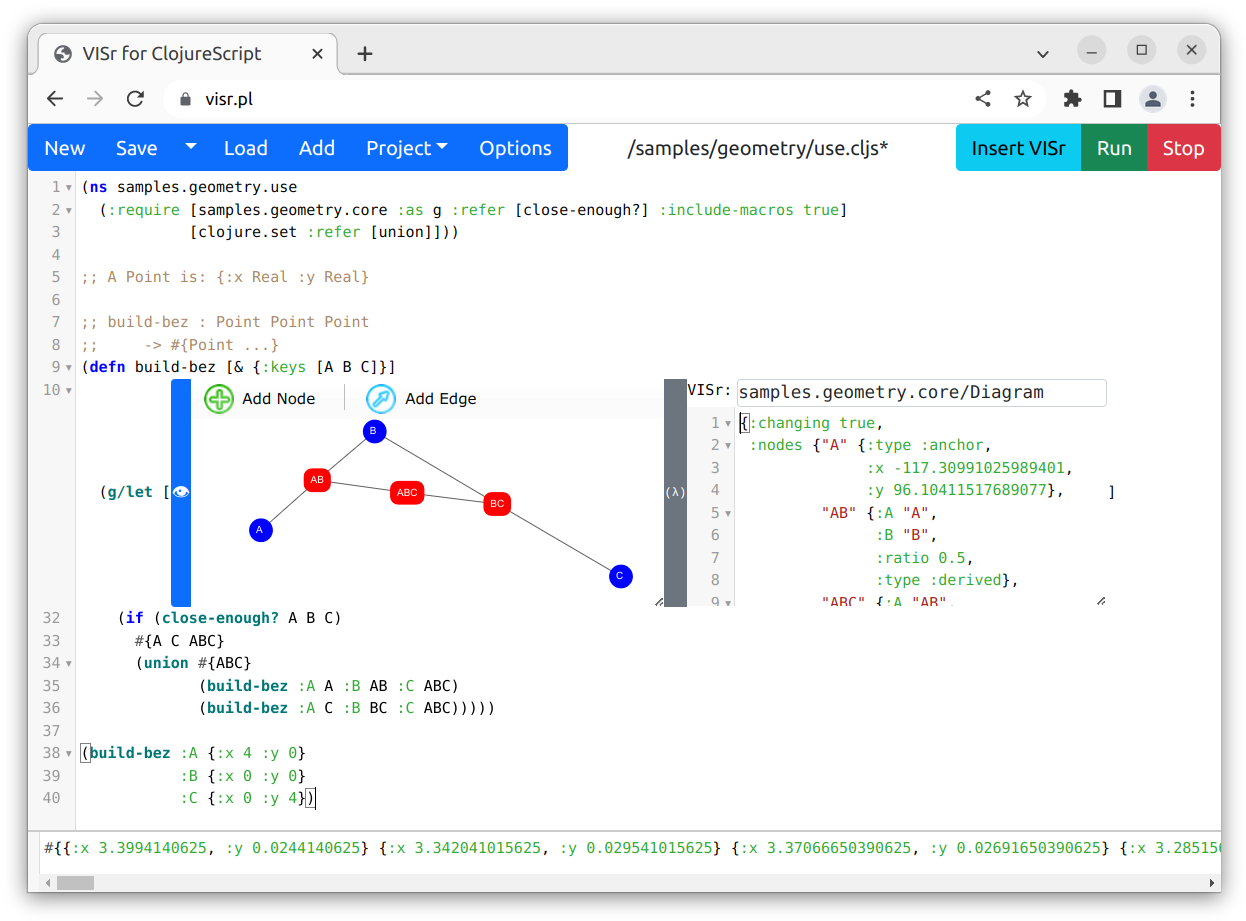}
  \makebox[0pt][r]{
    \raisebox{2em}{

    }\hspace*{1em}
  }
  \caption{In-IDE view of the Bézier function}
  \label{fig:bez-example}
\end{figure}

Figure~\ref{fig:bez-example} displays what a programmer sees when programming in
\textsc{Hybrid Clojure\-Script} using \texttt{elIDE}. Concretely, the screenshot shows the code of the
Bézier curve function from figure~\ref{fig:bezier} in \texttt{elIDE}. The most
interesting part of the screenshot is the use of the interactive-syntax
extension described in the introduction.

As the screenshot shows, the implementation of visual syntax for the
Bézier function renders the code in two ways: a visual view on the left and a plain
textual one on the right. The visual view generalizes the standard diagrams for
midpoint calculations that students might see in a geometry class. The diagram is
to be understood abstractly, meaning it computes midpoints based on the run-time
position of the given nodes (\texttt{A}, \texttt{B}, and \texttt{C}) and their relative position to each other.

Equally important is the purely textual representation of the code seen right next
to the visual view. This text is what the IDE puts into a file when the programmer
saves the code. Hence any ClojureScript implementation can run this saved version
of hybrid code. Better still, text is what IDE tools or command-line tools
process, which implies that programmers continue to benefit from all these tools
as they develop in the hybrid language. Finally, the text is also what programmers
see when they open the code in unadapted IDEs or plain-text editors.

Another important aspect of interactive syntax is that extensions implement a
model-view-control pattern. That is, a change to either view is immediately
reflected in the other one. A model---dubbed the state---reconciles the two views
with each other. When a programmer uses gestures to manipulate the mini-GUI, the
implementation of interactive syntax changes the state; the IDE notices the
change and updates both views.

\begin{figure}[ht]
  \includegraphics[scale=0.8]{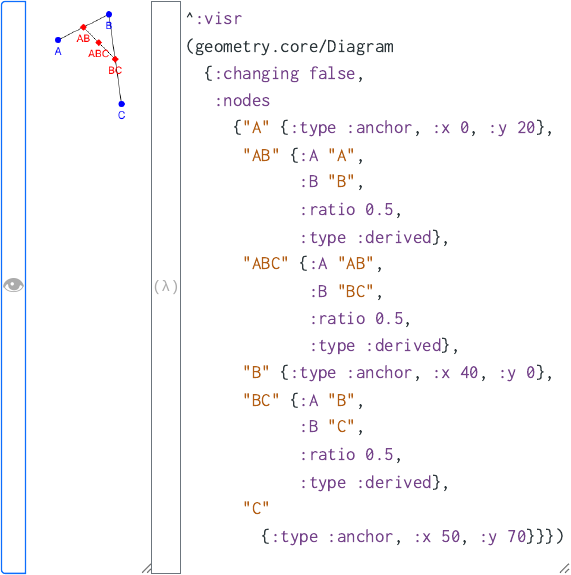}
  \caption{A close look at the Bézier-specific syntax extension}
  \label{fig:graph-view}
\end{figure}

Figure \ref{fig:graph-view} provides a close look at the plain-text view of an
interactive-syntax construct, which is just a function application.
Specifically, it applies a function from the definition of the
interactive-syntax extension to a textual version of the current state. Here the
reference is to the {\tt Diagram} extension, found in the {\tt geometry.core}
module.

This examination of the textual view demonstrates two points. First, a programmer
can create an instance of an interactive-syntax extension in a plain text
editor. No special IDE is needed. Second, a programmer can change the reference
pointer to the interactive-syntax definition ({\tt Diagram}), and the visualization on the left
would change immediately. (If \texttt{elIDE} cannot find the implementation,
it falls back on a default view.)

Even this brief tour validates how \textsc{Hybrid Clojure\-Script} running in \texttt{elIDE} satisfies
all desiderata of section~\ref{sec:design-space}.
Table~\ref{fig:design-comparison} contrasts the system with the work of
\citet{abf:adding}, the closest competitor. It clarifies how the system presented
here cooperates with IDE tools properly and preserves the existing
workflow. Further, the edit-time GUIs use the same library as run-time GUIs and
the IDE itself. Besides making the language and IDE easier to use,
it also makes \textsc{Hybrid Clojure\-Script} running in \texttt{elIDE} have far better
performance characteristics than Andersen et al.'s adaptation of Racket and
DrRacket.

\begin{figure}[htb]

  \def\foo{\hbox{Add Interactive Syntax to Existing, Textual PL}}

  \newdimen\stringwidth
  \setbox0=\foo
  \stringwidth=\wd0

  \def\prp#1{\hbox to \stringwidth{#1}}
  \def\row#1{\hbox to \stringwidth{#1\hfil}}

  \def\comp#1{\multicolumn{1}{c}{#1}}

  \def\yes#1{\hspace{.8cm}\cmark$^{#1}$}
  \def\noo#1{\hspace{.8cm}\xmark$^{#1}$}
  \def\yeah{\yes\relax}
  \def\nope{\noo\relax}

   \begin{tabular}{lll}
      \toprule
      \prp{Property \ \hfil Related Research} & \comp{\citet{abf:adding}} & \comp{this paper} \\
      \prp{\hfil Language}                    & \comp{Racket}             & \comp{ClojureScript}\\
      \prp{\hfil GUI library}                 & \comp{bespoke GUI}        & \comp{DOM} \\
      \midrule
      \row\foo                                             & \yeah         & \yeah \\
      \row{Adapt Popular, General-Purpose IDE}             & \noo*         & \yeah \\
      \row{Hybrid PL is backwards compatible}              & \yeah   	   & \yeah \\
      \row{Hybrid IDE is backwards compatible}             & \noo\dagger   & \yeah \\
      \row{Standard GUI library, GUI Component Reuse}      & \noo\ddagger  & \yeah \\
      \row{Linguistic visual and interactive syntax}       & \yeah   	   & \yeah \\
      \bottomrule
   \end{tabular}

~\\
~\\


  \begin{tabular}{r l}
    ${}^*$ & Limited to its hybrid capabilities when used in an IDE.\\
    ${}^\dagger$ & Standard IDEs are forwards compatible, only hybrid IDEs break compatibility.\\
    ${}^\ddagger$ & Some reuse is possible by using a shim to generate GUIs from a common source. \\
  \end{tabular}
  \caption{Desiderata comparison for interactive syntax designs}
  \label{fig:design-comparison}
\end{figure}

\section{A Recipe for Adapting a Language} \label{sec:design-lang}

If an existing programming language comes with bindings for an appropriate GUI
library, then turning it into a hybrid one can happen in a step-by-step fashion.
This section explains those steps, illustrates them with
ClojureScript, and presents a complete example in the hybrid variant.

\subsection{The Recipe}

A hybrid language allows programmers to add new, problem-specific syntactic
constructs to the already-available vocabulary. Programmers can then use these
constructs to build libraries or full programs with interactive and visual
syntax.

Given this context, the first step of the recipe requires creating syntax for
defining new kinds of interactive syntax. More specifically, this new definition
form specifies how interactive-syntax extensions keep track of their {\em
state\/} (of the model), i.e., those values that must persist; how they {\em
render\/} this state as a mini-GUI, or {\em serialize\/} it as plain text; and
how these mini-GUIs {\em react\/} to programmer gestures that, in turn,
manipulate the state. This setup closely follows the MVC architecture. Finally,
when the programmer wishes to run programs constructed with interactive visual
syntax, the interactive-syntax extension must know how to {\em elaborate\/} the
textual view to a run-time semantics; this may happen via a ``compilation''
or an ``interpretation''.


The visible novelty here is that the state of interactive-syntax extensions can
be rendered as either a mini-GUI for use in adapted IDEs or plain program text.
The latter is used in both adapted and non-adapted IDEs, as well
as when the entire program is run. Additionally, when the programmer interacts
with the code in either modality, the interactive-syntax state must be updated
so that both views show the up-to-date rendering as needed.

\begin{figure}[h]
  \centering
  \includegraphics[width=4in]{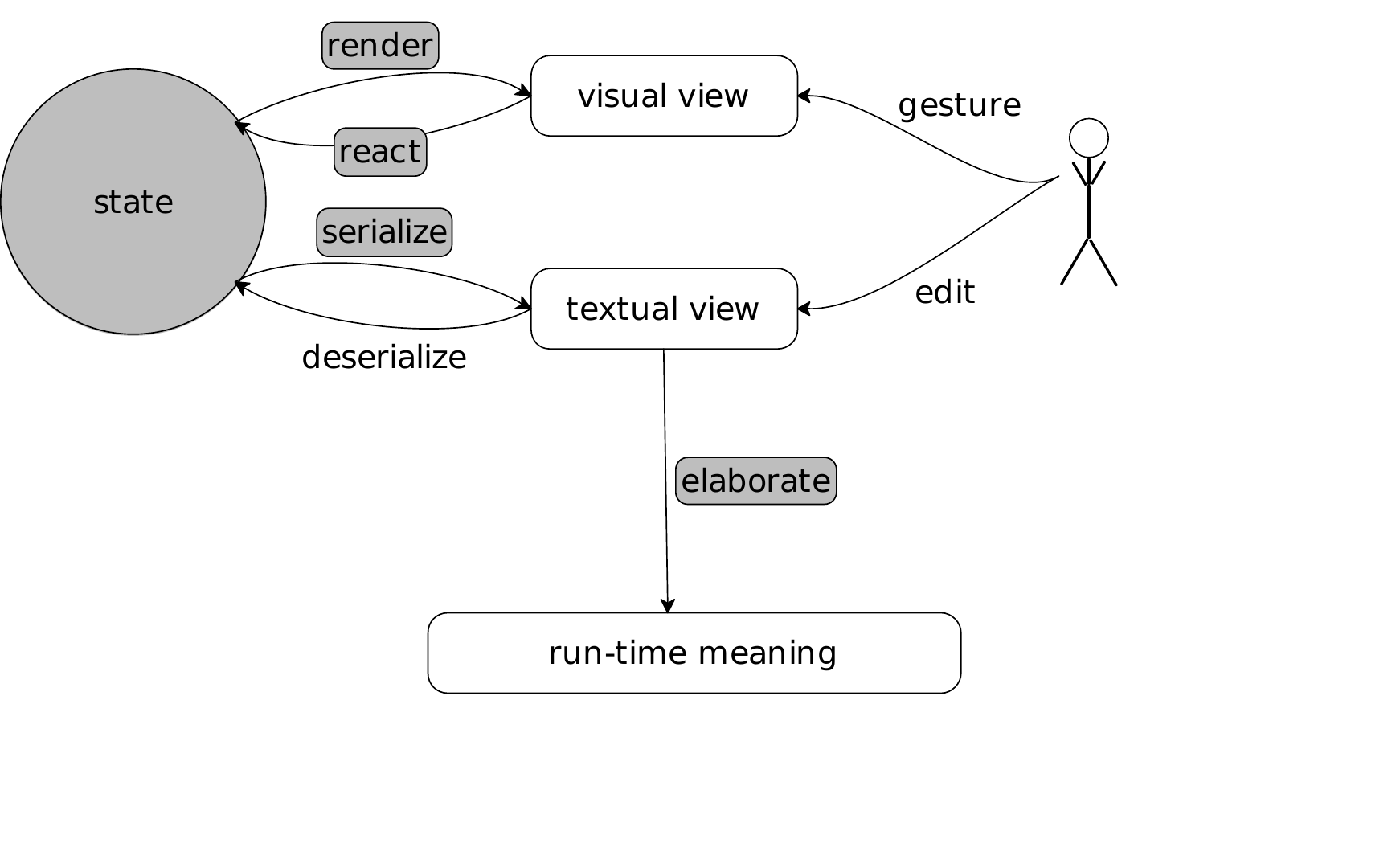}
  \vspace{-1cm}
  \caption{Interactive-syntax extensions at work} \label{fig:design}
\end{figure}

Figure~\ref{fig:design} sketches how an interactive syntax instance responds to
programmer stimuli. It highlights (in gray) the five elements that the creator
of an interactive-syntax extension must specify with the definition form.

The challenge for hybrid-language implementers concerns program phases. While
programming always involves three phases---{\em edit time\/}, when the
programmer edits the code; {\em compile time\/}, when the code is compiled; and
{\em run time\/}, when the resulting target code runs---interactive-syntax
extensions demand that {\em programmer-defined\/} code can run at edit time and
compile time.


For a language implementer to follow this recipe means picking a representation
for instances of an interactive-syntax extension---ranging from strings (bad) to algebraic
data types (acceptable) to S-expressions (fantastic, due to its synergy with the
multi-phase reflective nature of interactive-syntax extensions)---and to supply
an interpreter or a compiler for this notation. In the context of JavaScript,
for example, a ``cook'' could use a transpiler framework to assign semantics
to new syntactic elements. In the context of a macro-extensible language,
however, the work is even simpler; it suffices to implement a single (but
non-trivial) macro definition. The next subsection illustrates this particular
technique by applying the recipe to ClojureScript.

\subsection{Applying the Recipe to ClojureScript}

In order to turn ClojureScript into a hybrid language, it suffices to define a
single macro, named {\tt defvisr}, whose purpose is to define new interactive
syntax extensions. To use {\tt defvisr}, a programmer must specify three
elements:
\begin{enumerate}

\item a state element, which is an association of field names with initial values;

\item \texttt{render}, which equips the extension with edit-time semantics; and

\item \texttt{elaborate}, which assigns run-time semantics to the current state.

\end{enumerate}

Here is a template of the new definition form:

\medskip
\includegraphics[scale=0.8]{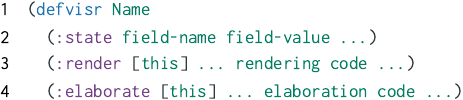}
\medskip

\noindent{}A \texttt{defvisr} definition introduces a new
interactive-syntax construct, which can be instantiated many times,
via plain text code or via GUI gestures (in a hybrid IDE). The
\texttt{state} component specifies the state part of the interactive-syntax
extension, as indicated by the ``state'' box in
figure~\ref{fig:design}.

The use of \texttt{defvisr} specifies three computations. First, \texttt{render}
consumes one argument---named \texttt{this} by convention---which is the current
state. It turns the state into a DOM element that is sent to the IDE, assuming
it is suitably adapted. Following standard DOM-development practice,
\texttt{render} collapses view and control. That is, it is simultaneously
responsible for drawing the GUI and for handling user input that allows the
direct manipulation of the state. For the second aspect, \texttt{render} may
mutate the fields of the state. In ClojureScript terminology, the \texttt{defvisr}
macro implementation supplies \texttt{render} with an ``atom'' containing the
state. Thus, the \texttt{render} component implements the ``render'' and ``react''
boxes from figure~\ref{fig:design}.

Second, the extension provides serialization for states using
JavaScript's serialization facilities, as required by the ``serialize''
box in figure~\ref{fig:design}. Observe, however, that a
\texttt{defvisr} definition does not require specifying serialization
explicitly. Instead, a \texttt{defvisr} instance implements this functionality
implicitly for the programmer.

Third, like \texttt{render}, \texttt{elaborate} consumes the current state (as text)
as its sole argument. Its task is to interpret the serialized state when the
ClojureScript program runs, as indicated by the ``elaborate'' box in
figure~\ref{fig:design}. The textual view expresses this idea with a call to the
elaborator (which actually has the same name as the interactive-syntax construct
itself, e.g., the \texttt{Diagram} \texttt{defvisr} defined below) wrapped around
the serialized state. This expression may end up being a function application
or, since this is ClojureScript, a macro. In the latter case, \texttt{elaborate}
may generate compile-time code, which can, for example, set up new variable
bindings or statically check instances of the interactive-syntax extension.

As a convenience, \texttt{defvisr} exploits the state specification to simplify
the syntax of the rendering and elaboration code. Specifically, it implicitly
binds the names of the fields of the state in the scope of the two function
bodies for reference. Mutation must use ClojureScript's \texttt{atom} functionality.

\begin{figure}[ht]
\centering
\includegraphics[scale=0.8]{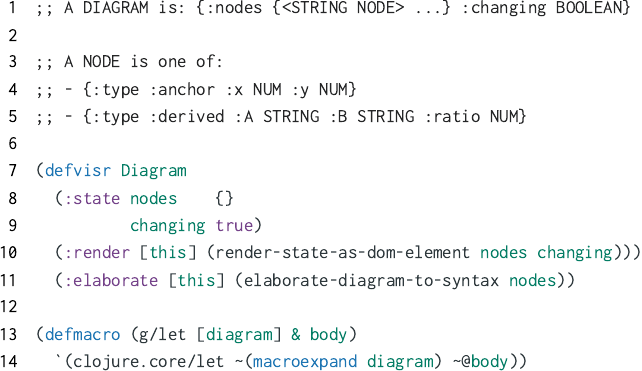}

\caption{The {\tt defvisr} for the interactive-syntax extension in figure~\ref{fig:bezier}}
\label{fig:the-defvisr}

\end{figure}

\subsection{Working with the Adapted ClojureScript}

Figure~\ref{fig:the-defvisr} sketches a \texttt{defvisr} definition of the
midpoint extension used in figure~\ref{fig:bezier}.

\paragraph{State} As the comments explain, the \texttt{Diagram} extension manages
values of type \texttt{DIAGRAM}. That is, the state consists of some \texttt{nodes}
and a boolean flag, called \texttt{changing}. Each \texttt{NODE} contains its type
and its position; the \texttt{nodes} field manages the \texttt{NODE}s and their
connections. The information in {\tt nodes} is used for drawing the diagram at
edit time as well as setting up variable bindings for the body of the plain-text
\texttt{g/let} macro at compile time. The \texttt{changing} field of the state is
set to true when the programmer is actively modifying the diagram; it is an
edit-time only value.

The \texttt{:type} field indicates that there are two distinct classes of nodes.
{\em Anchor\/} nodes are inputs to instances of the {\tt Diagram}
interactive-syntax extension; their positions become known at run time only. In
the example from figures~\ref{fig:bez-example} and~\ref{fig:graph-view},
\texttt{A}, \texttt{B}, and \texttt{C} are anchor nodes. {\em Derived\/} nodes are
outputs of the midpoint calculation; their values are determined algebraically
from anchor nodes and other derived nodes. In the previous example, \texttt{AB},
\texttt{BC}, and \texttt{ABC} are derived nodes. An interaction with the visual
diagram could shift these derived nodes and assign weights other than {\tt 0.5},
yielding a different kind of curve calculation.

\paragraph{The Renderer}

\begin{figure}[ht]
   \begin{subfigure}[c]{.5\textwidth}
     \centering
     \includegraphics[scale=0.8]{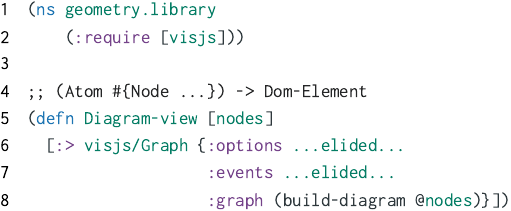}
     \caption{\texttt{library.cljs}}
   \end{subfigure}%
   \begin{subfigure}[c]{.5\textwidth}
     \centering
     \includegraphics[scale=0.8]{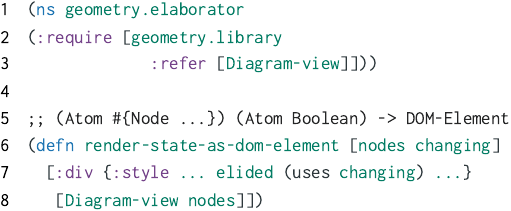}
     \caption{\texttt{renderer.cljs}}
   \end{subfigure}
   \caption{Renderer for a geometry extension}
   \label{fig:cljs-bez-render}
 \end{figure}

Figure~\ref{fig:cljs-bez-render} sketches the renderer implementation for
 the {\tt Diagram} interactive-syntax extension.
 Concretely, the code on the left side of the figure shows the function,
 \texttt{Diagram-view},  which renders a \texttt{Diagram} as a GUI view.
 It reuses functionality from a runtime GUI library and
 is thus straightforward for creators of the interactive syntax to write.
 More specifically, an
 external JavaScript library, \texttt{visjs}, handles the low-level drawing and event handling
 for the \texttt{Diagram}. The \texttt{:>} (line~6 on the left) is a special keyword that
 acts as a foreign function interface (FFI) to external JavaScript libraries.

\begin{figure}[ht]
\end{figure}


The code on the right side is the functionality needed to use
 this library code for the actual {\tt Diagram} renderer. The
 {\tt require} specification imports the library, in particular, the {\tt
 Diagram-view} function.
As mentioned, the {\tt render-state-as-dom-element} function is applied to an atom that
contains the state. An atom in ClojureScript is essentially a mutable box. From this state, the
renderer computes a data structure that encodes the user-facing DOM-element.
Functions placed in the first position in a vector (e.g. \texttt{Diagram-view} on
line~8) are treated as sub-components to be rendered. Likewise,
keywords (e.g. \texttt{:div} on line~7) directly represent DOM
tags.

This rendering code is also called in response to a programmer's interaction
with the mini-GUI. It then reads and modifies the state through interactions
with the state atom. Unboxing the atom, through the \texttt{@} operator (on the left),
returns an immutable encoding of the state (line~8, left).
When the state atom changes, a publish–subscribe style watcher (in the \texttt{visjs} library)
notices and updates the two views.

\begin{figure}[ht]
  \centering
  \begin{subfigure}[c]{0.35\textwidth}
    \includegraphics[scale=0.8]{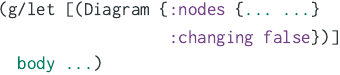}
    \end{subfigure}%
    \begin{subfigure}[c]{0.2\textwidth}
      \centering
      $\Longrightarrow$ \\
      elaborates
    \end{subfigure}%
    \begin{subfigure}[c]{0.45\textwidth}
      \includegraphics[scale=0.8]{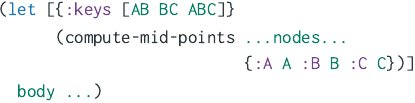}
  \end{subfigure}
  \caption{Elaborator for a geometry extension}
  \label{fig:cljs-bez-elab}
\end{figure}

\paragraph{The Elaborator}

Generally speaking, an interactive-syntax elaborator is a
syntax-to-syntax function. Figure~\ref{fig:cljs-bez-elab} shows an
example of how \texttt{Diagram}'s elaborator (which is invoked by
directly applying the \texttt{Diagram} name to a state representation) is used.
As seen on the left-hand side, this particular
\texttt{Diagram} interactive-syntax is designed to be used with a
special {\tt g/let} macro.  Specifically, elaboration of a
\texttt{Diagram} produces two parts: a sequence of identifiers (called
\texttt{keys} in ClojureScript), and an expression.

The right side of the figure shows how these elaboration results are
used in the run-time representation of the program: the three computed
identifiers are used as binders in a plain {\tt let}, and the
expression computes the values for these binders. More specifically,
the expression is an application of \texttt{compute-mid-points}---a
run-time function---to symbolic names and the concrete anchor nodes;
it computes the derived node positions based on the run-time position
of the anchor nodes.
All of this sets up the three variable bindings that are seen as the
three red midpoint dots in the visual interactive syntax from
figures~\ref{fig:bezier} and~\ref{fig:bez-example}.

\section{Adapting an IDE to \textsc{Hybrid Clojure\-Script}} \label{sec:design-ide}

A hybrid programming language must come with at least one IDE that can display
interactive-syntax extensions visually and textually.  This section describes
\texttt{elIDE}, an adaptation of CodeMirror for writing programs in
\textsc{Hybrid Clojure\-Script}. CodeMirror is a DOM-based editor that serves as the foundation of
a number of IDEs.


\subsection{\texttt{elIDE} Needs a Hint}

A closer look at figure~\ref{fig:graph-view} (on
page~\pageref{fig:graph-view}), specifically the textual view on the right,
reveals an explicit {\tt $\widehat{~}$:visr} Clojure metadata prefix
on the first line. This tag tells
\texttt{elIDE} that (1) this expression is an instance of interactive syntax and
(2) the expression contains a reference to the implementation for its
visualization. This use of metadata plays a key role in getting an adapted IDE
to work with hybrid syntax.

\subsection{The Architecture of \texttt{elIDE}} \label{sec:architecture}

Unlike an IDE for plain-text programming, an IDE for a hybrid language must run
{\em user-defined code\/} at edit time. Enabling an IDE to run user-defined code
at edit time raises a number of concerns, most importantly, that edit-time code
may interfere with the integrity of the IDE. Also, any design must be modular so
that the hybrid language and the hybrid IDE do not become tightly coupled.

\begin{figure}[h]
  \centering
  \includegraphics[width=0.8\textwidth]{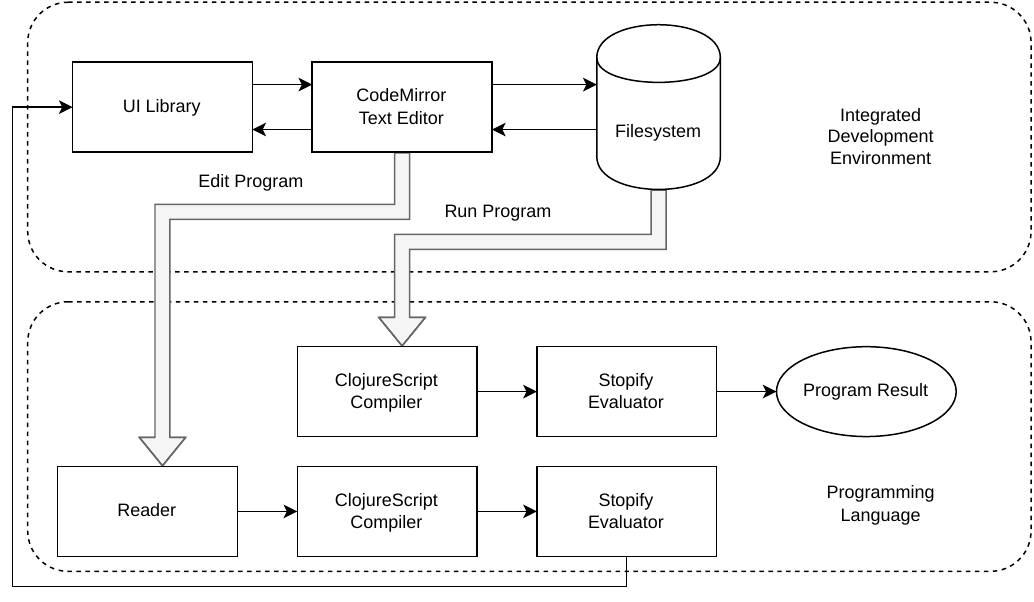}
  \caption{Architecture for \texttt{elIDE} (top) and \textsc{Hybrid Clojure\-Script} (bottom)}
  \label{fig:architecture}
\end{figure}

Figure~\ref{fig:architecture} shows an architecture diagram of \texttt{elIDE} (top)
and \textsc{Hybrid Clojure\-Script} (bottom). The rest of this subsection first explains the standard IDE
features and then those added for \textsc{Hybrid Clojure\-Script}.

\paragraph{Standard IDE Facilities} The \texttt{elIDE} IDE supports the expected
functionality:
(1) editing code with the CodeMirror text editor;
(2) storing code in a file and loading it from there;\footnote{Because
\texttt{elIDE} runs in a browser, it uses BrowserFS~\cite{pvb:browsix}, a
lightweight file system for browsers.}
(3) running code; and
(4) background execution.
All of these facilities are realized in the expected fashion.

\paragraph{Hybrid IDE Facilities}

The key hybrid facility is to reflect changes to the state of an instance of an
interactive-syntax extension in the program's displayed source code. Whether
these changes happen via text editing or direct manipulation of the GUI does not
matter. By allowing \textsc{Hybrid Clojure\-Script} to communicate back to \texttt{elIDE}, this
reflection is enabled in a natural manner.

Adding interactive syntax to a language means that a visual IDE must
continuously run the rendering code of instances of interactive syntax in the
background. As a programmer edits interactive syntax, the IDE updates the state
if needed and calls the appropriate functions to present the visualization and
the plain-text variants as needed.

A \textsc{Hybrid Clojure\-Script} specific reader (figure~\ref{fig:architecture}) is the entry
point for this task. It scans the entire program, determines which portion of
the code runs at edit time, and sends this portion of the code to the background
evaluator.

In order to satisfy the edit-time evaluation requirement, ClojureScript is
bootstrapped with an evaluator that composes the ClojureScript compiler and
Stopify~\cite{bnpkg:putting}. The latter is a JavaScript transpiler and run-time
environment whose purpose is to compile straight JavaScript into code that
supports cooperative multitasking through continuation passing.  In the context
of \textsc{Hybrid Clojure\-Script}, Stopify supplies two pieces of functionality. First, it allows
the IDE to pause running programs, i.e., misbehaving interactive-syntax
extensions do not lock up the IDE. Second, it provides a sandbox environment
that separates edit-time code for interactive-syntax extensions from the IDE. It
thus prevents the former from interfering with the IDE's internals.

Finally, the rendering code of interactive syntax is the only way for
\textsc{Hybrid Clojure\-Script} to send information back to the IDE. State managed in
this code is translated back into the program as text. Specifically,
the implementation turns the required changes into code snippets
written in a standard JavaScript library for manipulating the DOM. (The
right-hand side of figure~\ref{fig:cljs-bez-render} shows a concrete
example of what these snippets look like and how they are
computed.) When these pieces of code are sent to CodeMirror, they
place DOM elements into the text editor at the proper places.

\subsection{The General Idea}

The lessons learned from building a hybrid IDE with CodeMirror generalize to a
recipe. It starts from an IDE that uses the same GUI library, including an
editor, as the hybrid language. To make the IDE hybrid, that IDE must be able to
collaborate with hybrid languages so that it can display instances of
interactive syntax as mini GUIs. This collaboration covers three aspects:
\begin{enumerate}

\item the language implementation can tell the IDE which pieces of the code
are interactive;

\item the IDE can request that the language implementation evaluate the
GUI code from identified instances of interactive syntax at edit time and
complete programs at run time; and

\item the language implementation may insert elements into the IDE's editor.

\end{enumerate}
Without the last, the IDE cannot show programmers interactive syntax as GUIs
instead of text.

Hence, adapting an existing IDE to a hybrid language is easily implementable if
it comes with a plug-in API or a similar capability. This plug-in API must grant
full access to the IDE's editor, and it must support callbacks that allow the
IDE to invoke the language implementation on both pieces of functionality---at edit time---and
complete programs---at run time. The first kind of callback relies on the
above-mentioned full access capability so that it can insert the results of the
evaluation at the appropriate places. Finally, running the implementation at
edit time demands some form of sandboxing. For \textsc{Hybrid Clojure\-Script}, the authors had to manually construct this
sandboxing for the CodeMirror IDE using Stopify; if a team adapts another, more-powerful IDE,
such as Visual Studio Code,\footnote{https://code.visualstudio.com/} the existing language-server
architecture may already account for this need.\footnote{https://microsoft.github.io/language-server-protocol/}

\section{Evaluation: Preserving and Enhancing a Developer's Workflow} \label{sec:evaluation}

Evaluating a language design should confirm that it is both {\em useful\/} and
{\em usable\/}. Previous designs for interactive-syntax demonstrate its {\em
usefulness\/} with a plethora of examples. Each validates that interactive-syntax
expresses some domain concepts more directly and clearly than linear text. The
next section sketches how such examples are easily reconstructable in the
DOM-based approach.

The {\em usability\/} of previous designs, however, is questionable. Here,
usability means that developers can build on what they know and can easily create
and insert interactive-syntax extensions. More generally, a usable hybrid
language should enhance---not interfere with---the ordinary software development
workflow. This section presents a systematic characterization of major and minor
workflow activities and an analysis of how well this paper's DOM-based design
compares with prior designs to enhance and preserve them (section~\ref{sub:compare}).
Additionally, it addresses some remaining areas of \textsc{Hybrid Clojure\-Script} and \texttt{elIDE}
that need improvement (section~\ref{sub:flaws}).


\subsection{Workflow Operations and Interactive Syntax} \label{sub:compare}

Programmers interact with their codebases in the following major ways:

\begin{itemize}

\item \emph{Auditing}, the most common task, is reading and comprehending existing
  code. The primary goal of visual and interactive syntax is to let code about
  geometric concepts speak for itself.

\item \emph{Creation}, the second-most common task, is to write new code. As far
 as interactive syntax is concerned, ``creation'' refers to two actions: (1)
 creating new interactive-syntax extensions and (2) using existing
 interactive-syntax extensions (from a library) to create programs. In the ideal
 case, a programmer working in a text-only IDE can still insert an instance of
 an interactive-syntax extension, and it must work correctly in a hybrid IDE.

\item \emph{Copy and Paste} is the act of copying code to, and pasting it from, the
 clipboard. It also refers to the direct action of dragging and dropping. Both
 are common, and interactive syntax must not get in the way of either.

\item \emph{Running} programs (in the IDE or otherwise) is a fundamental part of
 software development. Existing tools should work without changes, even if
 programs include instances of interactive-syntax extensions.

\item \emph{Search and Replace} is the act of finding code and, optionally,
  replacing it with new code. At a minimum, interactive syntax should not hinder
  these operations. Ideally, a developer should be able to search for graphical
  renderings of interactive syntax and/or replace existing code with graphical
  renderings of interactive syntax.

\end{itemize}

In addition to these five major actions on code, there is a significant number
of more minor ones: \emph{Abstraction},
\emph{Autocomplete},
\emph{Coaching},
\emph{Code Folding},
\emph{Comments},
\emph{Comparison},
\emph{Debugging},
\emph{Dependency Update},
\emph{Elimination},
\emph{Hyperlinking Definitions and Uses},
\emph{Merging},
\emph{Migration},
\emph{Multi-Cursor Editing},
\emph{Refactoring},
\emph{Reflow},
\emph{Styling},
\emph{Undo/Redo}.
To avoid an overly
long and tedious evaluation section, however, this section deals only with the
major actions; a comparison of the minor actions with the most closely related
approach can be found in the last section.


  \def\foo{\hbox{\footnotesize Creation (definition)}}
  \newdimen\stringwidth
  \setbox0=\foo
  \stringwidth=\wd0

  \def\prp#1{\hbox to \stringwidth{\footnotesize #1}}
  \def\action#1{\row{\footnotesize #1}}
  \def\row#1{\hbox to \stringwidth{#1\hfil}}

  \def\comp#1{\multicolumn{1}{c}{\footnotesize #1}}

  \def\yes#1{\hspace{1cm}\cmark$^{#1}$}
  \def\noo#1{\hspace{1cm}\xmark$^{#1}$}
  \def\yeah{\yes\relax}
  \def\nope{\noo\relax}

\begin{figure}[ht]
  \begin{tabular}{lllll}
    \toprule
    \prp{Activity \hfil Literature}   & \comp{this paper}     & \comp{\citet{abf:adding}} & \comp{\cite{ocmvc:livelits}}          & \comp{\cite{sandblocks}} \\
    \prp{\hfil System}                & \comp{---}            & \comp{Hybrid Racket}      & \comp{Livelits}                       & \comp{Sandblocks}        \\
    \prp{\hfil Language}              & \comp{ClojureScript}  & \comp{Racket}             & \comp{bespoke language}               & \comp{Squeak}            \\
    \prp{\hfil GUI}                   & \comp{DOM}            & \comp{bespoke GUI}        & \comp{DOM}                            & \comp{Morphic}           \\
    \midrule
    \action{Auditing}               & \yeah                  & \yeah                    & \yeah                                 & \noo\ddagger \\
    \action\foo                     & \yes*                  & \yes*                    & \yes*                                 & \yes* \\
    \action{Creation (use)}         & \yeah                  & \noo\dagger              & \noo\dagger                           & \yeah  \\
    \action{Copy and Paste}         & \yeah                  & \yeah                    & \nope                                 & \yeah  \\
    \action{Running}                & \yeah                  & \yeah                    & \yeah                                 & \yes\S \\
    \action{Search and Replace}     & \yeah                  & \nope                    & \nope                                 & \yeah  \\
    \bottomrule
  \end{tabular}
  \begin{tabular}{r l}
    ${}^*$ & Orthogonal to interactive-syntax extensions.\\
    ${}^\dagger$ & Possible, but difficult. See Section~\ref{sec:design-space}.\\
    ${}^\ddagger$ & Sandblocks limits interactive-syntax components to expressions. \\
    ${}^\S$ & While possible, dynamic scoping leads to unexpected behavior. See Section~\ref{sec:related}.\\
  \end{tabular}
  \caption{Interactive syntax vs coding actions}
  \label{fig:edit-comparison}
\end{figure}


Figure~\ref{fig:edit-comparison} presents a comparison of the DOM-based design
and other systems with respect to the major workflow operations. Specifically, it compares with \citet{abf:adding}, Livelits~\cite{ocmvc:livelits} and
Sandblocks~\cite{sandblocks}, which are the closely related projects with similar goals.
Each cell in the two columns marks a workflow operation with a
``\cmark'' or a ``\xmark'', depending on how well the design works with this
action.




As mentioned, the always-available text view is the key reason why all
operations are possible with a DOM-based interactive-syntax approach. Every
instance of an interactive-syntax extension is always available as both a
graphical widget and a plain-text rendering because serialization works {\em
inside\/} the IDE. Furthermore, all instances of interactive-syntax extensions
are serialized to files as text, annotated with metadata. Hence every workflow
operation can exploit the plain-text version, both inside and outside of the
IDE.

The bespoke GUI library used in~\citet{abf:adding} is the key reason why two
major workflow operations are difficult to impossible in that system.
Specifically, the chosen IDE must internally store the bespoke GUI code as
binary data---rendering existing workflow operations unavailable. For
``creation,'' the bespoke GUI library does make it possible to create new types
of interactive-syntax extensions, but, using those new extension types is
extremely challenging. Developers must frequently leave the IDE entirely to make
relatively simple changes. Also, ``Search and Replace'' is limited to the
functionality of the bespoke GUI library. As a result, developers must once
again leave the IDE and use a plain-text editor.


Livelits uses a bespoke programming language. As such, the system fails to
support acquired programming habits. It does, however, use the DOM GUI system,
and it thus is conceivable that it could incorporate some ideas from this paper
in order to improve some developer workflow operations. While both the DOM-based
design of this paper and Sandblocks support all common editing operations,
Sandblocks modifies Squeak's lexical scoping to dynamic scope. Further,
interactive-syntax in this system may be created for expressions only. Thus,
Sandblocks users are limited in which parts of a program may be expressed
visually.

\subsection{Minor Limitations} \label{sub:flaws}

\textsc{Hybrid Clojure\-Script}, the hybrid language in this paper, is not perfect when it comes to
usability. This subsection describes three shortcomings, so that potential hybrid
language implementers are aware of them. None of these shortcomings are
fundamental to interactive-syntax extensions, however, nor are they fundamental
to a DOM-based design. Rather, they are limitations due to the chosen programming
language, ClojureScript.

First, ClojureScript's macro system requires putting macro definitions and uses
in separate files. Thus \textsc{Hybrid Clojure\-Script} introduces some workflow friction for
programmers who wish to develop extensions and test instances in a single file.
Fortunately, interactive-syntax extension definitions and uses can be placed in
one file if they compile directly to a single run-time function.

Second, ClojureScript's macro system implements only a weak form of
hygiene~\cite{kffd:hygiene, cr:mtw}. Thus, interactive-syntax elaborators are not
hygienic. To circumvent this weakness, \textsc{Hybrid Clojure\-Script} provides a functional version
of {\tt elaborate} which suffices in most cases.


Finally, as briefly mentioned in section~\ref{sec:design-ide}, \textsc{Hybrid Clojure\-Script} falls
short in its sandboxing capabilities. While this has not posed any problem for
any of the (almost 100) users of the prototype, it does highlight the need for
future research on the interface between interactive syntax and security. As is,
the limited sandbox provided by Stopify means interactive-syntax extensions are,
at worst, as secure as ordinary web pages designed with security in mind.

Although these limitations are undesirable, none of them reduce the usefulness or
usability of \textsc{Hybrid Clojure\-Script} in a substantial way. In practice, the ability to use a
rendering engine with multiple decades of engineering offsets the high friction
of defining extensions and minor problems with hygiene and sandboxing.

\section{Evaluation: Dom-Based Interactive Syntax At Work} \label{sec:examples}

A hybrid language based on a standard and widely used GUI library, i.e., a
DOM-based one, comes with significant advantages. First, programmers have built
many specialized libraries for multi-dimensional domains that can be used to
implement interactive-syntax extensions. Second, using such libraries imposes
almost no effort on a programmer. Third, if a library does not quite fit, it
tends to be open source and thus easy to modify.

This section presents three uses of \textsc{Hybrid Clojure\-Script} and implicitly \texttt{elIDE}.
The first one shows how to use a graph library as-is to express a REST API
connection as a state-machine. The second illustrates the ease with which a
slightly modified library can be used to express the geometry of a board game.
The final re-creates the most sophisticated example of~\citet{abf:adding}'s
work---a meta-syntax extension---with less code by reusing existing libraries.

\subsection{Using and Combining Existing Libraries} \label{sub:rest}

A protocol for calling methods in a certain order, e.g., during
authentication, is often expressed as a state-machine diagram. Thus it
is a perfect use case for interactive-syntax.
More specifically, to prevent misusing objects, library authors often inject run-time code which
checks that methods are called in the right order. Without interactive-syntax, this code must be manually synced
to the state-machine diagram present in the documentation. Of course, there is
no guarantee that the code corresponds to the diagram. Worse, the diagram and
the code are likely to get out of sync as the library evolves.

Using an interactive-syntax extension, a library author can describe a protocol
graphically and have the corresponding run-time-checking code generated
automatically. Consider an authentication protocol for a REST API on
objects with three methods: \texttt{auth}, \texttt{req}, \texttt{done}.
The protocol imposes the following constraints on these methods:

\begin{enumerate}
  \item use the \texttt{auth} method to send credentials and receive an
    authentication token in response;
  \item use the \texttt{req} method, with an endpoint URL and the valid authentication token,
    to repeatedly request data; and
  \item use the \texttt{done} method to end the authenticated session.
\end{enumerate}

\begin{figure}[ht]
  \centering
  \includegraphics[scale=0.8]{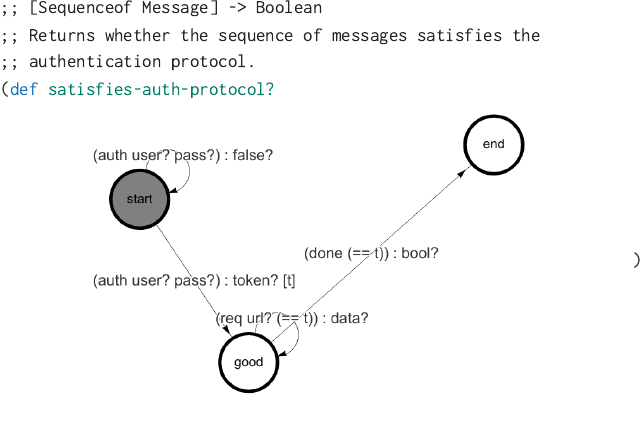}
  \caption{A state machine for an API protocol}
  \label{fig:automaton-graph}
\end{figure}

Figure~\ref{fig:automaton-graph} shows how a programmer may use an
interactive-syntax extension to express the protocol as a state machine. This
extension elaborates to a predicate which, given a sequence representing the
history of method calls to an authentication object, determines whether it
satisfies the protocol~\cite{dimoulas-pearl}.

The state machine consists of three states: {\tt start}, {\tt good}, and {\tt
end}. The shaded gray background of \texttt{start} shows that it is the starting
state. Each state indicates, via the transitions emanating from it, the set of
methods a client module can call. For example, in the \texttt{good} state, a
client module can call either the \texttt{req} or \texttt{done} method. A
transition is labeled with a method name plus predicates for the arguments and
result. If the arguments and result satisfy the predicates specified on the
transition, then the state machine moves to the next state. If no such
transition exists, then the protocol is violated, and this violation is reported.

In figure~\ref{fig:automaton-graph} the transition corresponding to a successful
authentication binds the returned token to the variable \texttt{t}. This is
shown in square brackets. The scope of this binding includes all downstream
transitions. Any transition in scope can then use this variable in predicates.
For example, the expression \texttt{(== t)} constructs a predicate that determines
if a value is equal to the token.

\begin{figure}[ht]
  \centering
  \includegraphics[scale=0.8]{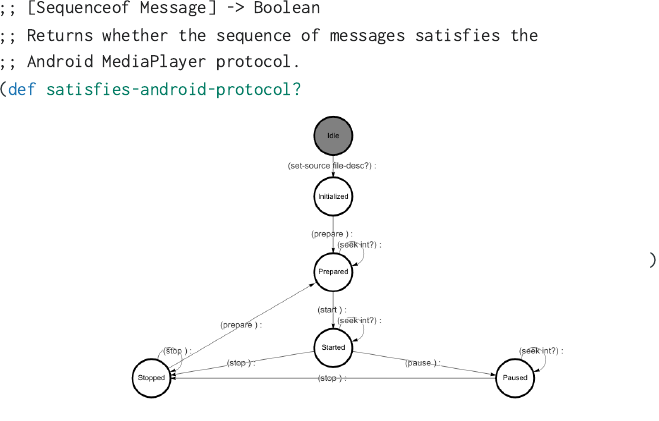}
  \caption{The Android MediaPlayer protocol with interactive syntax}
  \label{fig:android}
\end{figure}

The diagram presented in figure~\ref{fig:automaton-graph} is actually just one
use of a general-purpose interactive-syntax extension, which is used here to
generate state-machine-checking predicates. To demonstrate the versatility of
this extension, figure~\ref{fig:android} shows a slightly simplified
implementation of the Android MediaPlayer API\footnote{https://developer.android.com/reference/android/media/MediaPlayer}
protocol that uses the same interactive-syntax extension.

Using the interactive-syntax extension, a programmer performs GUI gestures to
create new states; delete existing ones; add or delete transitions; edit the
source and destination of a transition; turn states into starting or accepting
states; rename states (via a text box); edit the predicates labeling a
transition (via a text box); and change what variables are bound. These gestures
are intuitive. For example, creating a new transition merely requires clicking
and dragging from the source state to the destination state. Altering the
properties of a transition involves selecting the transition and clicking the
edit button.

The extension's elaborator analyzes code on the transitions to determine the
necessary binding structure. Specifically, the elaborator creates a separate
function for each transition with the appropriate parameters, and provides the
run-time system enough information to supply the correct arguments to each
function. Syntax and type errors in the specification are raised at compile
time. For example, if a transition predicate specified a dependency on a
variable that is not in scope, \texttt{elaborate} would signal a compile-time
error.

Developing this kind of interactive-syntax extension is a relatively low-effort
project. In a sense, it is a variation of the Bézier curve example as it uses the
same generic graph-drawing component.\footnote{https://visjs.org/} For the dialog to edit
transition edges, the extension depends on a different GUI
library.\footnote{\label{note:bootstrap}https://getbootstrap.com/} As a result of this
reuse-and-combine approach, the implementation for this extension consists of
fewer than 300 lines of code.
\subsection{Forking Libraries} \label{sub:catan}

Implementing a board game is another scenario where domain-specific geometric
ideas dominate a number of activities. This subsection examines how the popular
{\em Settlers of Catan\/} game can benefit from graphical syntax.

More specifically, implementing {\em Settlers\/} is challenging due to its
hexagonal grid board where each edge of a hexagon is colored according to the
player that owns that edge. A {\em road\/} consists of a continuous sequence of
edges of the same color. When the game is scored, the longest such road plays a
role.

\begin{figure}[ht]
  \begin{subfigure}[c]{.52\textwidth}
    \centering
     \includegraphics[scale=0.8]{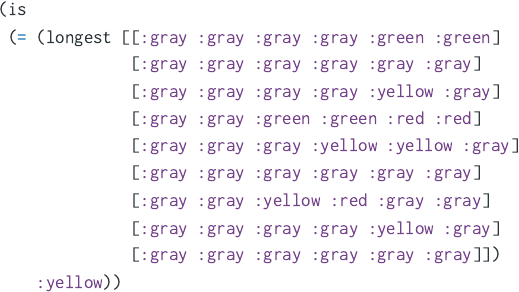}
    \caption{Textual unit test}
    \label{fig:catan-text}
  \end{subfigure}%
  \begin{subfigure}[c]{.48\textwidth}
    \centering
    \includegraphics[scale=0.8]{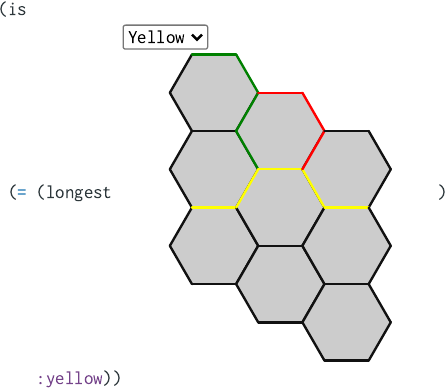}
    \caption{Visual unit test}
    \label{fig:catan-hybrid}
  \end{subfigure}
  \caption{An interactive-syntax extension for an implementation of ``Settlers of Catan''}
  \label{fig:catan}
\end{figure}

Unit tests demonstrate the usefulness of interactive-syntax extensions
particularly well. In this spirit, figure~\ref{fig:catan-text} displays a unit
test for the longest road calculation using traditional plain-text syntax. By
contrast, figure~\ref{fig:catan-hybrid} presents the same unit test using an
instance of interactive-syntax extensions for tiles and boards. The board shows
up exactly as it does in the application's GUI itself. Indeed, the
interactive-syntax extension reuses GUI code from the application itself, making
it simple to implement. If the GUI code of the game application were to change,
the syntax extension would tag along. One consequence of this reuse concerns the
manipulation of the board and tiles. To update the board, a programmer clicks
directly on an edge to change its color. The desired color is selected via a
drop-down menu (upper left).


It should be obvious from the two side-by-side figures that (1) text is an
inferior medium to express and maintain the unit test and (2) an
interactive-syntax representation comes with additional compile-time advantages,
such as well-formed test inputs and outputs.

Most importantly, \textsc{Hybrid Clojure\-Script} enables a programmer to implement the above
scenario with about 50 lines of code and a small adaptation to an open-source
library. Technically speaking, the code for the board uses a hexagon-grid
library,\footnote{https://github.com/Hellenic/react-hexgrid} a mostly generic component for drawing
hexagons. Unlike the libraries used in the preceding subsection, this library is
not extensible. Thus, the authors had to fork it and add 45 lines of code in
about two hours; these lines are generic, however, and would enhance the
existing library for many other purposes.
\subsection{Meta-Extensions} \label{sub:meta}

Meta-extensions are the most complicated form of interactive-syntax definitions.
Roughly speaking, a meta-extension is an interactive-syntax
extension whose instances elaborate to another interactive-syntax extension.

\begin{figure}[ht]
  \begin{subfigure}[b]{.5\textwidth}
    \centering
    \includegraphics[scale=0.8]{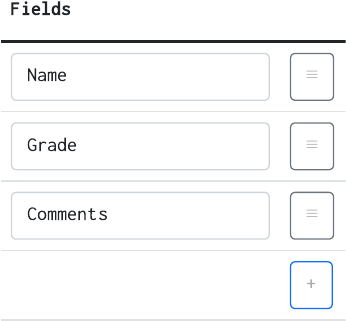}
    \caption{An assignment-specific form definition for graders}
    \label{fig:form-builder-def}
\end{subfigure}%
\begin{subfigure}[b]{.5\textwidth}
  \centering
  \includegraphics[scale=0.8]{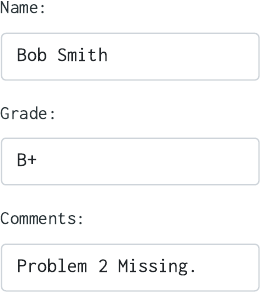}
  \caption{Instance of the assignment-specific form}
  \label{fig:form-builder-use}
\end{subfigure}
  \caption{An interactive-syntax extension for a form builder}
  \label{fig:form-builder}
\end{figure}

This example concerns the editing of (tabular) forms, which are useful in the
domain of software itself and many application domains. The case of editing
forms is obviously self-referential, meaning an editor for a form must be able
to generate forms. The concrete use case is about a programming instructor, who
makes grading forms that teaching assistants can use to report a student's
score.

Figure~\ref{fig:form-builder} illustrates how \textsc{Hybrid Clojure\-Script} can realize
such form editors.  Specifically, Figure~\ref{fig:form-builder-def} displays a
form editor for creating grading forms. In addition to creating new fields, the
instructor can reorder fields and add optional constraints on the data stored in
those fields. The {\tt elaborate} computation of this interactive-syntax extension
creates extensions whose instances look like the forms in
figure~\ref{fig:form-builder-use}. Once filled with student-specific data, these
generated forms elaborate to dictionaries, which can be submitted to the
instructor's gradebook code.

Both the form builder itself, as well as the forms created with that builder,
exploit another DOM-based GUI library.\textsuperscript{\ref{note:bootstrap}}
Using this library, the implementation is less than 50 lines of
code. For comparison,~\citet{abf:adding}, who created this example,
report that their form builder code comes
in at slightly more than 100 lines of code.

\section{Related Work} \label{sec:related}

The first subsection
compares with Sandblocks~\cite{sandblocks}, a system with similar characteristics to the one
presented in this paper. The remaining subsections look at works in other
several areas that inspired this research: (1) languages
and environments that allow programmers to run custom programs as they
edit code; (2) graphical and non-textual programming languages; and
(3) projectional and bidirectional editing.

\subsection{Sandblocks}
\label{subsec:sandblocks}

In February 2020, a research group at the Hasso-Plattner-Institut für Digitales
Engineering at Universität Potsdam published a technical
report~\cite{sandblocks} on the
Sandblocks system.
the Smalltalk programming language and its Morphic graphical development
environment. At first glance, the visual syntax extensions look related to the
ones presented here. Its visual extensions can be interleaved with program text.
Furthermore, the project report carefully spells out the design goal that visual
syntax must not interfere with a developer's tool chain and workflow.

Unfortunately, the design falls short of its goals. Unlike the hybrid
language presented in this paper, visual elements in the Sandblocks
implementation are not general. For example, programmers cannot add
visualizations for field definitions, methods, patterns,
templates, and other syntactic forms. Further, the visual constructs
do not respect the language's static semantics such as lexical
scope. As a result, the developer's toolchain and workflow are not preserved.



\subsection{Edit Time}

Two rather distinct pieces of work combine edit-time computation with a form of
programming. The first is found in the context of the Spoofax language workbench
project and is truly about general-purpose programming languages. The second is
Microsoft's mixing of textual and graphical ``programs'' in its Office
productivity suite.

Spoofax~\cite{kv:spoofax} is a framework for developing programming languages.
\citet{ekrkov:growing} recognize that, when developers grow programming
languages, they would also like to grow their IDE support. For example, a new
language feature may require a new static analysis or refactoring
transformations, and these tools should cooperate with the language's IDE. They
therefore propose a framework for creating edit-time libraries. In essence, such
libraries would connect the language implementation with the IDE and
specifically the IDE tool suite. The features are
extra-linguistic, however, and thus do not support the kinds of abstraction (and meta-abstraction) enabled by interactive-syntax extensions.

Microsoft Office plugins, called VSTO Add-ins~\cite{m:office}, allow authors to
create new types of documents and embed them into other documents. One developer
might use it to make a music type-setting editor, while another might use it to put music
notation in a PowerPoint presentation. Even though this tool set lives in the
.NET framework, however, it is also an extra-linguistic idea and does not allow developers to
build programming abstractions.

\subsection{Graphical and Live Languages}

Several programming \emph{systems} have enabled a mixture of some graphical and
textual programming for decades. The four most prominent examples are Boxer,
Hypercard, Scratch, and Smalltalk.

Boxer~\cite{da:boxer} allows developers to embed
GUI elements within other GUI elements (``boxing''), to name such GUI elements,
and to refer to these names in program code. That is, ``programs'' consist of
graphical renderings of GUI objects and program text (inside the boxes). For
example, a Boxer programmer could create a box that contains an image of a board
game tile, name it, and refer to this name in a unit test in a surrounding box.
Boxer does \emph{not}, however,  satisfy any of the other desiderata listed in
section~\ref{sec:design-space}. In particular, it has poor support for creating
new abstractions with regard to the GUI elements.

Scratch~\cite{rmmrebmrssk:scratch} is a fully graphical language system
widely used in education. In Scratch, users write their programs by
snapping graphical blocks together. These blocks resemble puzzle pieces and
snapping them together creates syntactically valid programs. Scratch offers
limited, but growing, capabilities for a programmer to make new block
types~\cite{hm:bringing}. These created block types, however, are themselves
created through text.

LabVIEW~\cite{k:labview} is a commercial visual language targeted at scientists and
engineers. It is widely adopted in its target communities. While it is
possible to create robust products using LabVIEW, extending it with new types of
visualizations is non-trivial, and it is rarely done.

Hypercard~\cite{g:complete} gives users a graphical interface to make
interactive documents. Authors have used Hypercard to create everything from
user interfaces to adventure games. While Hypercard has been used in a wide
variety of domains, it is not a general-purpose language.

Before the Sandbox project, Smalltalk~\cite{gr:smalltalk,iputm:lively,
bcdl:deep, kebmb:webstrates, rnaek:codestrates} supported direct manipulation
of GUI objects, often called live programming. Rather than separating code from
objects, Smalltalk programs exist in a shared environment called the
Morphic user interface~\cite{mrw:introduction}. Programmers can visualize GUI
objects, inspect and modify their code component, and re-connect them to the
program. No conventional Smalltalk system, however, truly accommodates general-purpose
graphical-oriented programming as a primary mode.

GRAIL~\cite{ehs:grail,ehs:grail-project} is possibly one of the oldest examples
of graphical syntax. It allows users to create and program with graphical flow
charts. Despite the apparent limitations of this domain, GRAIL was powerful
enough to be implemented using itself.

Notebooks~\cite{pg:ipython, a:observable, w:mathematica, bcdghhlmmmov:maple} and
Webstrates~\cite{kebmb:webstrates, rnaek:codestrates} are essentially a modern
reincarnation of GRAIL, except that they use a read-eval-print loop approach to
data manipulation rather than the GUI-based one made so attractive by the
Morphic framework. These systems do not permit domain-specific syntax
extensions.

\subsection{Bidirectional and Projectional Editing}

Bidirectional editors attempt to present two editable views for a program that
developers can manipulate in lockstep. One example,
Sketch-n-Sketch~\cite{chsa:programmatic, hllc:deuce}, allows programmers to
create SVG-like pictures both programmatically with text and by directly
manipulating the picture. Another example is Dreamweaver~\cite{a:dreamweaver},
which allows authors to create web pages directly and drop down to HTML when
needed. Changes made in one view propagate back to the other, keeping them in
sync. The interactive-syntax mechanism in this paper is more general, however, and thus the authors of this paper conjecture that it could be used to implement a bidirectional editing system. Dually, ideas from
other bidirectional editing systems could be used to improve the process of creating
interactive-syntax extensions in the future.

Wizards and code completion tools, such as Graphite~\cite{oylm:active}, perform
this task in one direction. A small graphical UI can generate textual code for a
programmer. However, once finished, the programmer cannot return to the
graphical UI from text.

Projectional editing aims to give programmers the ability to edit programs
visually.\footnote{Intentional Software~\cite{scc:intentional} seems related,
but there is little information in the literature about this
project.} Indeed, in this world, there are no programs per se, only graphically
presented abstract syntax trees (AST) that a developer can edit and manipulate.
The system can then render the ASTs as conventional program text. The most
well-known system is MPS~\cite{psv:jetbrains, vl:supporting}. It has been used
to create large non-textual programming systems~\cite{vrsk:mbeddr}. Unlike
interactive-syntax extensions, projectional editors must be modified in their
host editors and always demand separated edit-time and run-time modules. Such a
separation means all editors must be attached to a program project, they cannot
be constructed locally within a file. It therefore is rather difficult to
abstract over them.

Barista~\cite{km:barista} is a framework that lets programmers mix
textual and visual programs. The graphical extensions,
however, are tied to the Barista framework, rather than the
programs themselves. Like MPS, Barista saves the ASTs for a
program, rather than the raw text.

Larch~\cite{fkd:programs} also provides a hybrid visual-textual
programming interface. Programs written in this environment, however,
do not contain a plain text representation. As a result, programmers
cannot edit programs made in the Larch Environment in any other
editor.

The Hazel project and Livelits~\cite{ocmvc:livelits} are also closely related to
interactive-syntax extensions. Like editors, the Livelits proposal aims to let
programmers embed graphical syntax into their code. In contrast to
interactive-syntax extensions, which use phases to support editor instantiation
and manipulation, the proposed Livelits will employ typed-hole editing.

\citet{ek:expressive} introduced an Eclipse plugin that brought graphical
elements to Java. Like interactive-syntax, these graphical elements have a plain
text representation, stored as Java annotations. This implies that programmers
can write code with this plugin and view it in any plain-text editor. The plugin
differs from interactive-syntax extensions, however, in two ways: (1) the
plugins are less expressive than elaborators; and (2) the way new types of
extensions are created limits programmers' ability to abstract over them. For
example, programmers cannot create meta-instances with this plugin.

\section{Conclusion}\label{sec:recipe-final}

This paper describes a recipe for creating a hybrid programming language and IDE
which can support the exact right mix of textual and interactive visual code
that programmers need for their problem domain. Further, by starting with an
appropriate existing language, IDE, and GUI library, the recipe produces hybrid
results that are easy to use due to their familiarity to programmers; remain
compatible with unadapted language implemenations and tools; and also
preserve a programmer's workflow. Finally, the paper demonstrates these benefits
concretely by using the recipe to create \textsc{Hybrid Clojure\-Script} and a hybrid
CodeMirror-based IDE. The evaluation shows that they improve on many
of the shortcomings of prior hybrid textual-visual solutions.

The recipe discussed in this paper suggests several directions for future work.
Here are some examples of possible future directions. First, there are several
ways developers use visualizations when programming, everything from viewing
charts to displaying call graphs. A future study can classify these
visualizations and discuss how interactive syntax can handle them. Second, while
this paper provides a recipe for turning a language into a hybrid variant, a
language server protocol could be used to help automate this process. A future
attempt can describe and analyze this protocol. Finally, while the application of
the recipe to ClojureScript and CodeMirror is clearly successful, the next step
is to demonstrate the applicability of the recipe to a language that is not
already macro-extensible. The paper sketches how this can be accomplished; the
only true proof, though, is an actual implementation.


\section*{Acknowledgments}

This research was partially supported by NSF grants 1823244 and 20050550.

\clearpage
\appendix

\section*{Additional Workflow Evaluations}\label{sec:definitions}

Section~\ref{sec:evaluation} evaluated the usability of various hybrid language
systems with respect to the major operations that programmers use to
edit software. This section focuses on the more minor programmer actions.
Note that this evaluation considers a hybrid language usable even if it inhibits one
of the minor coding actions. Nonetheless, an analysis of the minor actions is still included,
with the understanding that each inhibited coding action increases the
friction programmers experience when using interactive syntax and decreases its usability.


  \def\same{\hspace{1cm}\textasciitilde}

  \def\foo{\hbox{{\footnotesize Hyperlinking Definitions and Uses}}}
  \newdimen\stringwidth
  \setbox0=\foo
  \stringwidth=\wd0

  \def\prp#1{\hbox to \stringwidth{\footnotesize #1}}
  \def\action#1{\row{\footnotesize #1}}
  \def\row#1{\hbox to \stringwidth{#1\hfil}}

  \def\comp#1{\multicolumn{1}{c}{\footnotesize #1}}

  \def\yes#1{\hspace{1cm}\cmark$^{#1}$}
  \def\noo#1{\hspace{1cm}\xmark$^{#1}$}
  \def\yeah{\yes\relax}
  \def\nope{\noo\relax}

\begin{figure}[ht]
  \begin{tabular}{lllll}
    \toprule
    \prp{Activity \hfil Literature}   & \comp{this paper}     & \comp{\citet{abf:adding}}  \\
    \prp{\hfil System}                & \comp{---}            & \comp{Hybrid Racket}       \\
    \prp{\hfil Language}              & \comp{ClojureScript}  & \comp{Racket}              \\
    \prp{\hfil GUI}                   & \comp{DOM}            & \comp{bespoke GUI}         \\
    \midrule
    \action{Abstraction}		         & \nope	         & \yeah \\
    \action{Autocomplete} 		         & \yeah 		 & \yeah \\
    \action{Coaching} 			         & \yeah		 & \yeah \\
    \action{Code Folding} 		         & \nope 		 & \yeah \\
    \action{Comments} 			         & \same 	         & \same \\
    \action{Comparison} 			 & \nope 		 & \yeah \\
    \action{Debugging} 			         & \nope 		 & \yeah \\
    \action{Dependency Update}      		 & \yeah 		 & \yeah \\
    \action{Elimination} 	   	     	 & \nope 		 & \yeah \\
    \action\foo 	 			 & \nope 		 & \yeah \\
    \action{Merging}                 	 	 & \nope      		 & \yeah \\
    \action{Migration} 			       	 & \same 		 & \same \\
    \action{Multi-Cursor Editing} 	 	 & \same 	  	 & \same \\
    \action{Refactoring}  		       	 & \nope 		 & \yeah \\
    \action{Reflow} 			         & \yeah 		 & \yeah \\
    \action{Style} 			         & \yeah 		 & \yeah \\
    \action{Undo/Redo} 			         & \yeah 		 & \yeah \\
    \bottomrule
  \end{tabular}
  \begin{tabular}{r l}
    \same & Orthogonal to interactive-syntax extensions.
  \end{tabular}
  \caption{Interactive syntax vs coding actions, continued (also see \ref{fig:edit-comparison})}
  \label{fig:edit-comparison-2}
\end{figure}

The minor programmer actions that are evaluated are:
\begin{itemize}
\item \emph{Abstraction} means generalizing two (or more) pieces of code into a
 single one that can then be instantiated to work in the original places (and
 more). Interactive syntax must facilitate converting one type of instance into
 another if abstraction involves the code for a definition of interactive
 syntax.

\item \emph{Autocomplete} allows programmers to choose descriptive names and
 enter them easily; recent forms of this code action complete entire phrases of
 code. It requires semantic knowledge of the programming language. In the ideal
 case, an IDE for a hybrid language should support autocompletion of textual
 prefixes into an instance of interactive syntax.

\item \emph{Coaching} is about the back-and-forth between programmers and
 analysis tools. A coaching tool displays the results of a (static or dynamic)
 analysis in the editor and (implicitly) requests a reaction. A simple example
 is the underlining of unbound variables; an advanced one may highlight expressions
 that force register spilling. The challenge is that adding interactive syntax
 means extending the language in a non-functional manner, and doing so comes
 with its inherent problems.

\item \emph{Code Folding} enables IDEs to hide blocks of code while editing.
 Developers like to present overviews of code with code folded. Interactive
 syntax must not inhibit this IDE action.

\item \emph{Comments} do not affect interactive syntax.

\item \emph{Comparison}, often referred to as diffing, is needed to comprehend
 small changes to existing code as those are created.  While the tool is
 available on say the text of git repositories, it is more often used inside of
 IDEs. If interactive syntax always comes with textual equivalents, code
 comparisons should continue to work in the conventional manner.

\item \emph{Debugging} demands running a program in a step-by-step fashion,
 i.e., steps a person can move through and comprehend sequentially. The comments of
 the preceding bullet apply here.

\item \emph{Dependency Update} is about updating packages and libraries for
 various reasons.  A change made to the definition of an interactive-syntax
 extension is reflected in uses of that extension automatically.

\item \emph{Elimination} is the dual of abstraction, meaning in-lining the code
 for an existing abstraction. The comments concerning the abstraction bullet
 apply here, too.

\item \emph{Hyperlinking Definitions and Uses} allows programmers to easily
 navigate between definitions and uses. To hyperlink
 pieces of code properly, an IDE must understand both the text and the
 semantics of code.  Whether this form of linking works properly depends on how
 easily an IDE can get hold of the text that corresponds to an instance of
 interactive syntax.

\item \emph{Merging} two blocks of code is the natural extension to
 comparison. Instead of viewing a report of the difference, however, merging
 attempts to generate syntactically correct code that represents two sources
 derived from one original point. Like comparison, merging is clearly a
 text-based action but more sophisticated. Research is needed to investigate how
 well merging works in the presence of interactive syntax.

\item \emph{Migration} happens when a dependency or platform changes, breaking
 backwards compatibility. Frequently this requires small tweaks through an
 entire codebase. Supporting extension migration suffices here.

\item \emph{Multi-Cursor Editing} allows two (or more) developers to
 concurrently edit the same program. This is orthogonal to interactive syntax.

\item \emph{Refactoring} is a syntax- or even semantics-aware search-and-replace
 action. Most simple refactoring actions should work as-is even in the presence
 of interactive syntax. More research is needed to understand whether
 refactoring works when syntactic differences involve interactive syntax.

\item \emph{Reflow} automatically transforms program text in an IDE buffer to
 conform to some style standards, e.g., proper indentation. If an IDE
 accommodates instances of interactive syntax, reflow continues to work.

\item \emph{Styling} changes aspects of code display, e.g., the font size or the
 color theme. Instances of interactive-syntax may benefit from explicitly
 coordinating with style operations.

\item \emph{Undo/Redo} is straightforward for text. For interactive syntax, each
 extension can package multiple changes into a single undo/redo step.
\end{itemize}

The table in figure~\ref{fig:edit-comparison-2} shows
how~\citet{abf:adding}'s design compares with the one presented in
this paper for each operation.

\clearpage
\bibliography{paper}
\bibliographystyle{plainnat}

\end{document}